\documentstyle[12pt,a4]{article}
\newcommand{\n}{\newline}
\begin{document}
\bibliographystyle{plain}
\baselineskip= 18pt
\input psfig
\input mssymb
{\Large \bf \noindent On the dynamics of a spherical spin-glass
in a magnetic field }
\vskip 2 truecm
\pagestyle{empty}
\vskip 0.5 truecm
\noindent {\bf L.F. Cugliandolo and  D.S. Dean}
\vskip 1 truecm
\noindent Service de Physique de L'Etat Condens\'e, Saclay, CEA,
\n
Orme des Merisiers, 91191 Gif--sur--Yvette Cedex,
\n
France.
\vskip 1 truecm
\noindent{\bf PACS:} 05.40, 05.70L, 75.40G.
\vskip 1 truecm \noindent{\bf Abstract:}
We carry out an analysis of the effect of a quenched
magnetic field on the dynamics of the spherical Sherrington-Kirkpatrick
spin-glass model.
We show that there is a  characteristic time
introduced by the presence of the field.
Firstly, for times sufficiently small the dynamic scenario
of the zero field case - with aging effects - is reproduced.
Secondly, for times larger
than the characteristic time one sees equilibrium dynamics.
This dynamical behaviour is reconciled with the geometry of
 the energy landscape of the model. We compare this behaviour with
experimental observations at a finite applied field.
\vskip 4 truecm
\noindent  29th May 1995
\newpage
\pagenumbering{arabic}
\pagestyle{plain}
\def\half{{1\over 2}}
\def\OO{\Omega}
\def\sech{{\rm sech}}
\def\n{{\newline}}
 \def\aa{\alpha}
 \def\bk{{\bf k}}
 \def\bkp{{\bf k'}}
 \def\bqp{{\bf q'}}
 \def\bq {{\bf q}}
 \def\EE{\Bbb E}
 \def\EEx{\Bbb E^x}
 \def\EEo{\Bbb E^0}
 \def\LL{\Lambda}
 \def\PP{\Bbb P^o}
 \def\rr{\rho}
 \def\SS{\Sigma}
 \def\ss{\sigma}
 \def\ll{\lambda}
 \def\dd{\delta}
 \def\ww{\omega}
 \def\ll{\lambda}
 \def\DD{\Delta}
 \def\DDt{\tilde {\Delta}}
 \def\kr{\kappa\lb \LL\rb}
 \def\PPx{\Bbb P^{x}}
 \def\gg{\gamma}
 \def\kk{\kappa}
 \def\tt{\theta}
 \def\bs{\hbox{{\bf s}}}
 \def\bh{\hbox{{\bf h}}}
 \def\lb{\left(}
 \def\rb{\right)}
 \def\prt{\tilde p}
\def\pt{\tilde {\phi}}
 \def\bb{\beta}
 \def\hal{{1\over 2}\nabla ^2}
 \def\bg{{\bf g}}
 \def\bx{{\bf x}}
 \def\bu{{\bf u}}
 \def\by{{\bf y}}
 \def\hag{{1\over 2}\nabla}
 \def\beq{\begin{equation}}
 \def\eeq{\end{equation}}
 \def\cosech{\hbox{cosech}}
\section{Introduction}
For some time there has been much interest in aging phenomena and
in general non-equilibrium behaviour \cite{Str}.
For a wide range of models one finds that
if the thermodynamic limit is taken before the long-time limit,
for quite
generic initial configurations of the system, one never sees an equilibrium
regime
characterised by time translational invariance of the correlation function and
the
satisfaction of the fluctuation dissipation theorem \cite{Cuku1,Frme}. Recently
the
authors presented a detailed examination of the dynamics of the spherical
Sherrington-Kirkpatrick
(SK) spin-glass model \cite{Kosetal} in Ref.\cite{Cude}, hereafter refered as
I.
In general the initial conditions do not lead the system to equilibrium and the
correlation function was shown to exhibit the
aging phenomenon. This model joins a whole range of models where
non-equilibrium
behaviour is present, however the precise physical mechanisms responsible for
aging
in various models can be completely different but still give the same
mathematical
behaviour for various dynamical functions. Aging can arise due to the existence
of
large energy barriers in the system (see {\it e.g.} Refs.\cite{Bo,Bode}), also
due to the
existence of zero modes in the free-energy landscape as in the case of the
models considered
in Refs.\cite{Cukupa,Cude},
due to a combined effect of large energy barriers and extended flat regions in
the energy landscape
as in mean-field models \cite{Cuku1, Frme}
and also via domain growth or coarsening mechanisms in the system \cite{Bray}.
In all these cases the correlation functions exhibit aging effects. However,
the
domain growth as well as the pure zero-mode mechanisms do not seem to be enough
to
capture aging in quantities such as the magnetisation. The effects of
small perturbations in such models are rapidly erased and henece the response
decays
 too fast to reproduce the slow relaxations of glassy systems.

In this paper we revisit the spherical SK model but
with the addition of a quenched  magnetic field.
For simplicity we work with a random magnetic field but this does not change
the
underlying physics of the problem.
The theory of linear response applied to this model would suggest that, for
sufficiently small fields, the behaviour of the system is not drastically
altered and that one should recover the type of aging phenomena observed in I.
However the random field strength appears
coupled to a {\it nonlinear} term in the equation for the dynamical Lagrange
multiplier
imposing the spherical constraint and hence it is not apparent that linear
response theory
should hold.
We shall concentate on the case of zero temperature where the discussion of
both the statics
and dynamics is most clear.
This is however sufficient to bring out the sentient points of the problem.
 One finds that the random magnetic field introduces a time scale into the
problem.
Below this characteristic time one indeed sees aging phenomena and the effects
 seen in I are reproduced.
Above this timescale the system does indeed {\it reach} an equilibrium state.
With the benefit of this insight we carry out a simple analysis of the energy
landscape and point out that it is the
disappearance of the preponderance of zero modes in the energy landscape
which is responsible for the ultimate equilibrium behaviour, adding
weight to the assertions made in I.
\section{Dynamics}
We shall recall briefly the definition of the model. The Hamiltonian is given
by
\beq
H = -{1\over 2}\sum_{ij} J_{ij} s_i s_j - \sum_i h_i s_i ,
\eeq
subject to the spherical constraint $\sum s_i ^2 = N$, (N being the number of
spins).
The matrix $J$ is a random symmetric matrix with independently Gaussianly
distributed
components scaled with $N$ to give the Wigner semi-circle law distribution for
the
eigenvalues $\mu$ in the thermodynamic limit, that is
\beq
 \rho(\mu) = {1\over 2 \pi} \sqrt{4 - \mu^2} \;,  \;\;\;\;\;\;\; \mu\in [-2,2].
\eeq
The terms $h_i$ represent the random field which is chosen to be Gaussian with
zero mean and
 variance $h^2$.
The dynamical equation for the evolution of the spins is then given by
 (at zero temperature) \cite{Cude, Cipa}
\beq
{\partial s_i \over \partial t} = J_{ij} s_j - z(t) s_i + h_i .
\eeq
We shall assume uniform initial conditions for the spins
\cite{Cude}
and proceed by diagonalizing the equations of motion and computing the
dynamical Lagrange multiplier $z$ self-consistently. Defining
\beq
\Omega(t)
\equiv
 \exp\lb \int_0 ^t z(t') dt' \rb
\eeq
we find it obeys the following equation
\beq
\Omega^2(t) = f(t) + h^2 \int_0^{t} dt'\int_0^{t} dt''
f\lb t-{t' + t'' \over 2} \rb \Omega(t')\Omega(t''),
\label{eq:dyn}
\eeq
where
\beq
f(t) \equiv \int \rho(\mu) d\mu e^{2\mu t} .
\eeq
In constrast with the zero field case the equation
satisfied by the dynamical Lagrange multiplier is now
 non-linear (where as before it was a linear Volterra
equation \cite{Cude}). The equation (\ref{eq:dyn}) is
evidently very difficult to solve, however it is clearly
causal and accessible to numerical solution. We procede by
making the following ansatz on the asymptotic form for $\Omega$
\beq
\Omega(t) \sim c \, e^{\ll t} ,
\eeq
where $c$ is some positive constant.
Futhermore if we assume {\em a priori} that $\ll > 2$
then because $f(t) \sim e^{4t}
(2t)^{-3\over 2}/\sqrt{4 \pi}$ for sufficiently large
times we may make the asymptotic approximation
\beq
\Omega^2(t) =  h^2 \int_0^{t} dt'\int_0^{t} dt''
 f\lb t-{t' + t'' \over 2 }\rb \Omega(t')\Omega(t'')
\; . \label{eq:dyn2}
\eeq
Assuming the $\Omega$ is well behaved and bounded
for small times we may also assume that the exponetial
behaviour dominates the double integral to obtain
the following equation determining $\ll$:
\beq
1 = h^2 \int d\mu{ \rho(\mu)\over (\ll -\mu)^2 }
 =
 {h^2\over 2}
\left(-1 + {\ll\over \sqrt{\ll^2 - 4}}
\right)
\label{eq:ll}.
\eeq
The equation (\ref{eq:ll}) yields the solution
\beq
\ll = {2 + h^2 \over \sqrt{1 + h^2}}, \eeq
hence we see {\em a posteriori} that indeed $\ll > 2$ for $h > 0$. Of course
the fact
that we have found an asymptotic solution for Eq.(\ref{eq:dyn}) does not mean
that
it is the solution that matches with the given initial conditions (the problem
of
matching in these kinds of  systems is in general outstanding, {\it e.g.} see
\cite{Cuku1}).
However we have checked numerically that this is indeed the case.
The results of a numerical integration of Eq.(\ref{eq:dyn})
for $h=0.5$ and $h = 0.4$ are shown in figures 1 and 2 respectively. The
function plotted is actually $\omega(t) = e^{-2t} \Omega(t)$. The figures
clearly show the onset of an
exponential behaviour of $\omega(t)$ around the characteristic time scale $\tau
(h)$, with the analytically predicted rate (plotted as the dashed line); also
shown is the zero field result (dotted line).

\par Hence we see via the above analysis that the addition of the random
magnetic
field introduces a time scale $\tau$ into the problem such that for times
sufficiently large
compared to $\tau$ the contribution of the first term of the righthandside of
Eq.(\ref{eq:dyn})
 becomes negligible compared to the second term. (It is in fact the first term
that contains
all the information about the initial condition; the fact that it becomes
negligible
demonstrates that the initial condition is completely forgotten at large enough
times.)
The analysis in addition shows that
\beq
\tau \sim 1/(\ll -2) = \lb {2 + h^2 \over \sqrt{1 + h^2}} -2\rb^{-1} ,
\eeq
which clearly diverges when $h\to 0$ as $h^{-2}$. However for sufficiently
small
fields and/or times it is the first term of the righthandside of
Eq.(\ref{eq:dyn}) which dominates, and this leads  to the aging phenomena
exhibited in \cite{Cude} within these field/time regimes.

\par The explicit demonstration of equilibrium behaviour for sufficiently
large times is now rather trivial. The correlation function is given by
\beq
C(t,t') = {1\over \Omega(t)\Omega(t')}
\left[
f\left({t+t'\over 2}\right) +
 h^2 \int_0^{t} ds\int_0^{t'} ds' f\lb {t + t' -s -s' \over 2} \rb
\Omega(s')\Omega(s)
\right]
\; .
\label{eq:CF}
\eeq
Inserting the asymptotic form for $\Omega$ simply yields
\beq
C(t,t') \sim 1
\; ,
\eeq
for sufficiently large times. Hence the system ultimately reaches its
equilibrium state,
 which is simply the point of minimal energy in the zero
temperature case.

\par The energy density ${\cal E }(t)$ can be shown to obey
\beq
{\cal E }(t) = -{1\over 2} {d\over dt} \log(\Omega) -{1\over 2}\aa(t),
\eeq
where $\aa$ is the average spin/field correlation
\beq
\aa(t) = {1\over N}\sum h_i s_i(t) = {h^2\over \Omega(t)}\int_0^t dt'\,
 f \left({t-t'\over 2} \right)\Omega(t')
\; .
\eeq
One finds that
\beq
{\cal E}(t) \to -\sqrt{1 + h^2} \label{eq:energy}
 \eeq
and
\beq \aa(t) \to {h^2\over \sqrt{1 + h^2}} \label{eq:alpha}.\eeq

\section{Zero Temperature Statics}

In this section we carry out a simple analysis of the geometry of the energy
landscape for the model in the presence of a random magnetic field. The energy
density corresponding to a spin configuration $\bs$ is given by
\beq
N{\cal E} = -{1\over 2}\bs\cdot J\bs - \bh\cdot \bs + {\ll\over 2}\lb
\bs\cdot\bs -N \rb ,
\eeq
where $\ll$ is the (static) Lagrange multiplier imposing the spherical
constraint. The variational equations  yield the stationary solutions
\beq
 \bs = (\ll -J)^{-1}\bh
 \eeq
where $\ll$ satisfy the equation
\beq
{1\over N} \; \bh (\ll -J)^{-2}\bh = h^2 \int {\rho(\mu) \over (\ll-\mu)^2}
d\mu
\; .
\eeq
For $\ll \in [-2,2]$ the integral above diverges, and hence for finite $h$
we must look for solutions outside this interval. There are only two and are
given by
\beq \ll_{\pm} = \pm {2 + h^2 \over \sqrt{1 + h^2}} .\eeq
The corresponding values of the energy density can be shown to be
\beq {\cal E}_{\pm} = -{1\over2} h^2 \int {\rho(\mu) \over (\ll_{\pm}-\mu)}d\mu
-{1\over 2} \ll_{\pm}. \eeq
The minimum of the energy is given by $\ll_{+}$, and $\ll_{-}$ gives the
maximum energy (which must also exist according to Morse theory). The minimum
energy and the static value of $\aa$ are in agreement with the dynamically
calculated limits in equations (\ref{eq:energy}) and (\ref{eq:alpha}).
\par If one analyses the zero magnetic field case one finds that there are $N$
stationary points for the energy, each one corresponding to an eigenvector of
the interaction matrix. Hence the effect of the field statically, has been to
reduce the number of stationary points from a macroscopic number down to just
two. (This is in sharp contrast with the reduction of the number of metastable
states
in the SK model by a magnetic field, where the number always remains
macroscopic
for finite fields \cite{De}.)
Therefore from almost every starting configuration (except the
maximum which is however measure zero), the system has a drift towards the
equilibrium state.
It was postulated  in \cite{Cude} that the  large number of flat regions in
the zero field landscape was responsible for the aging phenomena observed.
Here we see clearly that once these are erradicated an equilibrium state is
achieved.

\section{Conclusions}
We have analysed the influence of a  magnetic field
on the dynamics of the spherical SK spin-glass. At variance
with the prediction of linear response, the aging behiour of
 the system is completey erradicated after a characteristic time
which decreases with increasing field strength. The failure of linear
response is presumably due to the inherent nonlinearity introduced
in the equation of motion for the dynamical Lagrange multiplier.
However for sufficiently short times, linear response is presumably
valid and one does indeed see that below this time scale the aging
phenomenon is present. In fact
this model explicitly demonstrates the phenomenon of interrupted aging
discussed in \cite{France1}.
\par Also we have related the ultimate arrival at equilibrium of
the system to the geometry of the energy landscape of the model,
showing that the field reduces the number of stationary (flat) points
in the energy landscape from a macroscopic number to just two. This
is in agreement with the notion that in the spherical SK spin-glass
 in the absence of a random magnetic field, it is the zero modes
in the system which give rise to the aging phenomenon rather
than tunnelling through large energy barriers.

The behaviour of the spherical SK model is reminiscent of what
is predicted by the scaling approach to  spin-glasses
\cite{droplets}. Within the droplet picture,
the low-temperature phase is characterised by only
two pure states as for the spherical SK model.
The other hall-mark
of the droplet picture is the absence \cite{Mobr} of a genuine $h-T$
transition line (AT line) \cite{AT} and hence the destruction of the
thermodynamic spin-glass phase by any applied field.
However, a pseudo AT line defined as the locus of points at which the system
gets
frozen on experimental times scales moving down with increasing time
scales is claimed to exist in this approach  \cite{Mobr}.
For the spherical SK model, statically there
is no AT line but
dynamically we find that
the non-equilibrium aging behaviour exists below a transition line that moves
towards zero field when the times explored increase.

Experimentally there is no clear evidence for the appearance of an equilibrium
regime for spin-glasses \cite{Vi} under the influence of a magnetic field.
However, we expect that the behaviour observed
in this letter may be present in other disordered systems  such as the problem
of a driven particle in a random potential.

\vspace{1cm}

{\underline{ACKNOWLEDGEMENTS}}
We would like to thank J.-P. Bouchaud,
J. Kurchan, M. M\'ezard and E. Vincent for useful discussions. D. S. D.
acknowledges support from a UK EPSRC Research Fellowship and also from
EU grant CHRX-CT93-0411. L. F. C. acknowledges support from the
the EU HCM grant ERB4001GT933731.

\newpage
\pagestyle{empty}
\noindent{\bf List of Figure Captions}
\vskip 1 truecm
\noindent Figure:1 Graph of $\omega(t)$ verses time for $h = 0.5$.
\vskip 0.5 truecm
\noindent Figure:2 Graph of $\omega(t)$ verses time for $h = 0.4$
\newpage

\pagestyle{empty}
\begin{figure}[htb]
\centerline{\psfig{file=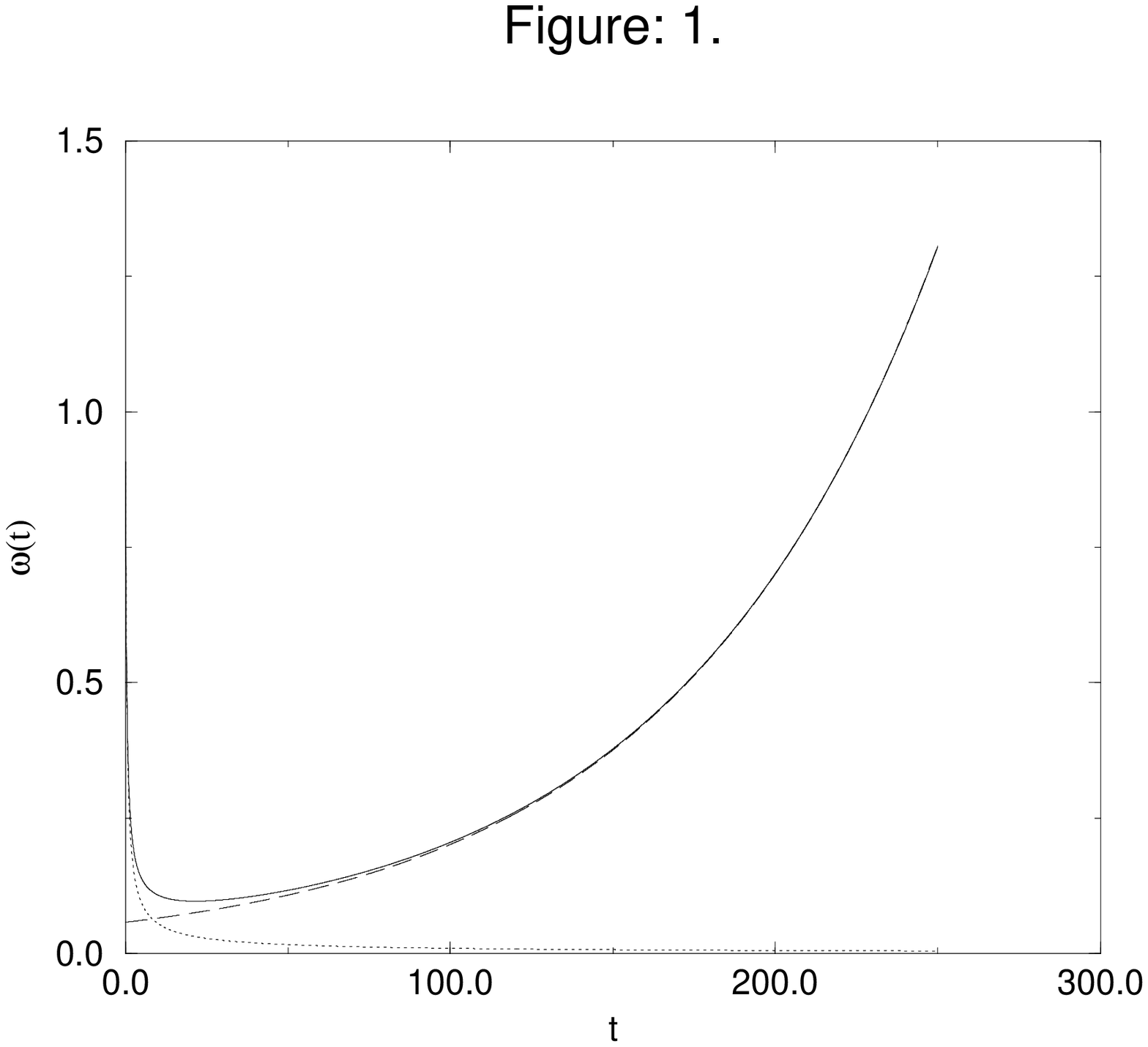}}
\end{figure}

\begin{figure}[htb]
\centerline{\psfig{file=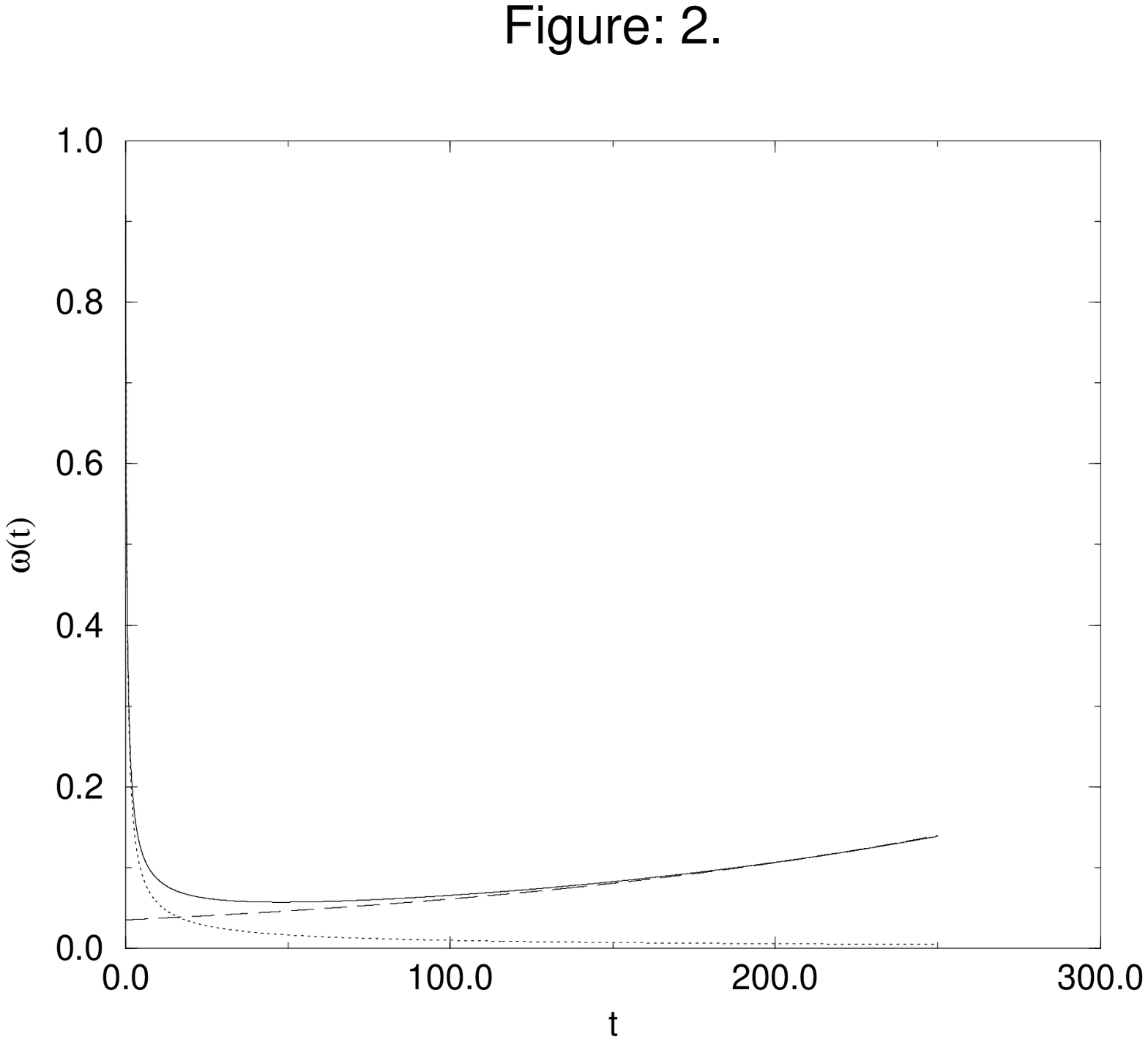}}
\end{figure}


\newpage
\baselineskip =18pt
\begin{thebibliography}{10}
\bibitem{Str}
L. C. E. Struik;
{\em Physical Aging in Amorphous Polymers and Other Materials}, Elsevier, North
Holland Inc., 1978.
\n
L. Lundgren, P. Svedlindh, P. Nordblad and  O. Beckman;
Phys. Rev. Lett.,  51:911 (1983).
\n
E. Vincent, J. Hammann and  M. Ocio; in {\it `Recent progress in
Random Magnets'}, ed. D. H. Ryan, World Scientific, Singapore (1992).



\bibitem{Kosetal}
J.M. Kosterlitz, D.J. Thouless and R.C. Jones
\newblock {\em Phys. Rev. Lett}, 36:1217, 1976.

\bibitem{Cude}
L.F. Cugliandolo and D.S. Dean
\newblock {\em Full dynamical solution of a mean-field spin-glass model},
cond-mat 9502075, to appear {\em J. Phys. A.}


\bibitem{Bo}
J.P. Bouchaud;
\newblock {\em J. Phys. I (France)}, 2:1705, 1992.

\bibitem{Bode}
J.P. Bouchaud and D.S. Dean
\newblock {\em J. Phys. I France}, 5:265, 1995.


\bibitem{Cukupa} L. F. Cugliandolo, J. Kurchan and G. Parisi;
\newblock {\em J. Phys. I (France)}, 4:1641 (1994).


\bibitem{Cuku1}
L.F. Cugliandolo and J. Kurchan
\newblock {\em Phys. Rev. Lett.}, 71:173, 1994,
{\em J. Phys. A.}, 27:5749, 1994.
\n
L. F. Cugliandolo and P. Le Doussal;
\newblock  {\em Dynamics of a particle in a random
potential}, cond-mat 9505112.

\bibitem{Frme}
 S. Franz and M. M\'ezard;
\newblock {\em Europhys. Lett.} 26:209 (1994),
{\em Physica A} 209:1 (1994).
%

\bibitem{Bray}
A. J. Bray
\newblock to appear in {\em Advances in Physics}.

\bibitem{Cipa}
S. Ciuchi and F. de Pasquale
\newblock {\em Nucl. Phys.} {\bf B300} [FS22], 31, 1988.


\bibitem{De}
D.S. Dean
\newblock {\em J. Phys. A}, 27:L899, 1995.

\bibitem{France1}
J.-P. Bouchaud, E. Vincent and J. Hammann
\newblock {\em J. Phys. I (France)}, 4:139, 1994.

\bibitem{Mobr}
A. J. Bray; {\em The ordered phase of a spin-glass}, {\em Comments Cond. Mat.
Phys.},
14:21, 1988.
\n
M. A. Moore; {\em Scaling theory of the ordered phase of real spin glasses},
in
{\em Cooperative dynamcis in complex physical systems},
Ed. H. Takayama, Springer-Verlag, 1989.

\bibitem{droplets}
A. J. Bray and M. A. Moore; Phys. Rev. Lett. 58:57, 1987.
\n
G. J. Koper and H. J. Hilhorst; J. Phys. (France) 49:429, 1988.
\n
D. S. Fisher and  D. Huse; Phys. Rev.  B 38:373, 1988.

\bibitem{AT}
J.R.T.de Almeida and D.J.Thouless; {\em J. Phys. A} 11:983 (1978).


\bibitem{Vi} E. Vincent; private communication.
\n
F. Lefloch, J. Hammann, M. Ocio and E. Vincent; {\em Physica B} 203:63, 1994

\end{thebibliography}
\end{document}